\documentclass[preprint,aps,showpacs]{revtex4}
\usepackage{epsfig}
\newcommand{\Be}{\begin{equation}}
\newcommand{\Ee}{\end{equation}}
\newcommand{\Bea}{\begin{eqnarray}}
\newcommand{\Eea}{\end{eqnarray}}
 
\newcommand{\NL}{\nonumber \\}                    %%
\newcommand{\mtxfnt}{\usefont{U}{dsss}{m}{n} \selectfont}  %useSpecialFont 
\begin{document}

\setcounter{page}{0}

\title{Large Mixing from Small: \\
Pseudo-Dirac Neutrinos and the Singular Seesaw}
\author{G.J. Stephenson, Jr.}\email{ GJS@baryon.phys.unm.edu}
\affiliation{Department of Physics and Astronomy, 
University of New Mexico, Albuquerque, NM 87131}
\author{T. Goldman}\email{ tgoldman@lanl.gov}
\affiliation{Theoretical Division, MS-B283, 
Los Alamos National Laboratory, Los Alamos, NM 87545}
\author{B.H.J. McKellar}\email
{ b.mckellar@physics.unimelb.edu.au} 
\author{M. Garbutt}\email{ mgarbutt@treasury.gov.au}
\affiliation{University of Melbourne 
Parkville, Victoria 3052, Australia}

\begin{flushright}
\vspace{-1.5in}
{LA-UR-04-1736}\\
\vspace{-0.1in}
{hep-ph/0404015}\\
\vspace*{0.2in}
\end{flushright}

\begin{abstract}

If the sterile neutrino mass matrix in an otherwise conventional 
seesaw model has a rank less than the number of flavors, it is 
possible to produce pseudo-Dirac neutrinos. In a two-flavor, 
sterile rank 1 case, we demonstrate analytic conditions for large 
active mixing induced by the existence of (and coupling to) the 
sterile neutrino components. For the three-flavor, rank 1 case, 
``3+2'' scenarios with large mixing also devolve naturally as we 
show by numerical examples.  We observe that, in this approach, 
small mass differences can develop naturally without any requirement 
that masses themselves are small.  Additionally, we show that 
significant three channel mixing and limited experimental resolution 
can combine to produce extracted two channel mixing parameters at 
variance with the actual values. 

\end{abstract}

\maketitle

\section{Introduction}

Conventional wisdom holds that neutrinos ought to be Majorana 
particles with very small masses, due to the action of a 
``seesaw'' mechanism\cite{see-saw}, which is built on the 
concept of quark-lepton symmetry\cite{GSMcK}.  Alternatively, 
there have been theoretical suggestions regarding pseudo-Dirac 
neutrinos in the past\cite{wolfm2,oldref}, and again more 
recently\cite{GSMcK,CHM,othr}, i.e., that neutrinos may well be 
Majorana particles occuring in nearly degenerate pairs. These 
can occur within the framework of the so-called ``singular'' 
see-saw where the rank of the mass matrix describing the 
(presumed to be) heavy neutrinos, which have no interactions 
(often referred to as ``sterile'' neutrinos) in the standard 
model~(SM), is less than maximal.

Recent results from Kamiokande\cite{superK} on atmospheric 
neutrinos, from Sudbury\cite{SNO} on solar neutrinos, and from 
KamLand\cite{KamL} on long baseline reactor neutrinos, appear 
to require oscillations between nearly maximally mixed (active 
neutrino) mass eigenstates.  Each of these analyses, however, 
argues that this mixing cannot be dominantly to sterile states 
such as are found in pseudo-Dirac pairs. On the other hand, the 
concatenation of the data from these experiments with that from 
LSND\cite{LSND} and other short baseline data does not appear 
to fit into a theoretical structure which only includes mixing 
among three active Majorana neutrinos.  Many have therefore 
been motivated to consider the effects of additional (sterile, 
Majorana) neutrino states, the existence of which is accepted 
in the conventional ``see-saw'' extension of the SM, although 
there the actual states are generally precluded from appearing 
directly in experiments by an assumption that the masses of the 
sterile states are very large. 

We investigate here how small flavor mixing effects in the 
sterile sector can lead to large mixing among active neutrinos 
in the presence of a singular see-saw. (In Ref.\cite{CHM}, 
large mixing  was achieved by means of a mass hierarchy in 
the Dirac mass sector.) Paralleling a convention in the quark 
sector, we assume the mass and flavor bases for the charged 
leptons are simultaneously diagonal, so that all flavor violations 
and oscillation phenomena are described as arising from the 
neutrino mixing angles alone. 

It should be noted that there is no accepted principle 
that specifies the flavor space structure of the mass 
matrix assumed for the sterile sector. Some early 
discussions\cite{see-saw,wolfm2} implicitly assume that 
a mass term in the sterile sector should be proportional 
to the unit matrix.  This has the pleasant prospect, 
in terms of the initial argument for the see-saw, that 
all active neutrino flavors have small masses on the 
scale of other fermions.  However, since there is no 
obvious requirement that Dirac masses in the neutral 
lepton sector are the same as Dirac masses in any 
other fermionic sector, this result is not compelling.  
Indeed, Goldhaber has argued for a view of family 
structure and self-energy based masses that naturally 
produces small neutrino masses\cite{Maurice}. We 
discuss here a more conventional possibility which 
arises from a minimal modification of the standard see-saw, 
namely that the rank of the mass matrix for the sterile 
sector is less than the number of flavors. Note that 
this does not conflict with quark-lepton symmetry which 
applies only to the number and character of states. 

In this paper, which is an extension of reference\cite{hep}, 
we shall concentrate on the case of a rank $1$ sterile matrix, 
relegating the rank $2$ case to some remarks at the end. (The 
analysis of short baseline data by Sorel, Conrad and 
Shaevitz\cite{Sorel} suggests that the rank $2$ case may not 
actually occur in Nature.) We note in passing that some Grand 
Unified Theories include more than $3$ fermions that are 
neutral under all of the interactions in the SM; a $4\times4$ 
or larger, rank $1$ sterile mass matrix could lead to $3$ 
pseudo-Dirac pairs of neutrinos involving all of the active 
neutrinos of the SM. 

Concentrating on a $3$-dimensional sterile space, we consider 
rank $1$ to be a natural case because whatever spontaneous 
symmetry breaking produces mass in that flavor space necessarily 
defines a specific direction. Before including the effects of the 
sterile mass, we assume three non-degenerate Dirac neutrinos, with 
Dirac masses, $m_1~<~m_2~<~m_3$, (although this is not essential,) 
which are each constructed from one Weyl spinor which is active 
under the $SU(2)_W$ of the SM and one Weyl spinor which is sterile 
under that interaction. (Being neutrinos, both Weyl fields have no 
interactions under the $SU(3)_C$ or the $U(1)$ of the SM.) There 
is then an MNS matrix\cite{MNS} which relates these Dirac mass 
eigenstates to the flavor eigenstates in a manner completely 
parallel to that of the CKM matrix\cite{ckm} for quarks.  Note, 
however, that these matrix elements are not the ones extracted 
directly from experiment, as the mass matrix in the sterile sector 
induces additional mixing.

We next use the Dirac mass ($m_{D}$) eigenstates to define 
bases in both the $3$-dimensional active flavor space and the 
$3$-dimensional sterile flavor space\cite{fn1}. Following the 
spirit of the original see-saw, we exclude any initial Majorana 
mass term in the active space. If the Majorana mass matrix in 
the sterile space were to vanish also, the three flavors of 
Dirac neutrinos would be a mixture of (Dirac) mass eigenstates 
in a structure entirely parallel to that of the quarks. 

A rank $1$ sterile mass matrix may be represented as a vector 
of length $M$ oriented in some direction in the $3$-dimensional 
sterile space. If that vector lies along one of the axes, then 
the Dirac neutrino that would have been formed from it and its 
active neutrino partner partake of the usual see-saw 
structure (one nearly sterile Majorana neutrino with mass 
approximately $M$ and one nearly purely active neutrino with 
mass approximately $m_D^2/M$) and the other two mass eigenstates 
remain Dirac neutrinos. If that vector lies in a plane 
perpendicular to one axis, the eigenstate associated with that 
axis will remain a pure Dirac neutrino, and the other two pairs 
of states form one pseudo-Dirac pair and a pair displaying the 
usual see-saw structure. Both of these pairs are mixtures of the 
$4$ Weyl fields associated with the two mixing Dirac neutrinos. 
In general, the structure consists of $2$ pseudo-Dirac pairs and 
one see-saw pair, all mixed.

As we implied above, the very large mixing required by the 
atmospheric neutrino measurements could have been taken to be 
evidence for a scheme involving pseudo-Dirac neutrinos. (This, 
after all, follows Pontecorvo's initial suggestion\cite{BP}.) 
However, pure mixing into the sterile sector is now strongly 
disfavored\cite{nsm}.  It is evident from the discussion above 
that there is a region of parameter space (directions of the 
vector) in which the two pseudo-Dirac pairs are very nearly
degenerate, giving rise to the possibility of strong mixing in 
the active sector coupled with strong mixing into the sterile 
sector.  We explore this point here. 

The organization of the remainder of the paper is as follows: 
In the next section we discuss a two flavor, $4\times 4$ neutrino 
mass matrix analytically. In Sec.\ref{genmass}, we present 
the general $6\times 6$ mass matrix and discuss the parameterization 
of the sterile mass matrix and various limiting cases.  We show 
the spectrum for a general case.  In Sec.\ref{2pds}, we apply 
our analysis to the case where the plane in question is 
perpendicular to the axis for the middle value ($m_2$) Dirac 
mass eigenstate, raising the possibility of near degeneracy  
between pseudo-Dirac pairs. Moving away from that plane produces 
large  mixing amongst the members of those pseudo-Dirac pairs. 
In Sec.\ref{example}, we show an example of the oscillation 
patterns that are produced and how limited experimental resolution 
can lead to errors in the extraction of physical parameters if 
the data analysis assumes only two channel mixing.  Finally, we 
remark on the structures expected for a rank $2$ sterile matrix 
and then reiterate our conclusions.

\section{Two flavor case}\label{2flavor}

In our examination of the consequences of assuming a rank $1$ 
mass matrix in the sterile subspace, we will show below that 
there are certain parameter ranges for which there is very 
large mixing induced in the active subspace, even though there 
is no explicit mixing among the original Dirac bispinors.  To 
see how this arises, it is useful to look at the two flavor 
model for which we can obtain an analytic description of the 
mass eigenvalues as a power series in $\frac{1}{M}$.  We then 
can find the eigenfunctions, again as a power series in
$\frac{1}{M}$, and look at the ratio of the coefficients for 
the two active components.  We examine the conditions which 
allow for large active mixing when there is no mixing in the 
original Dirac space.

In the next section we shall discuss the case where two 
pseudo-Dirac pairs are nearly degenerate and follow 
the mixing patterns as we move away from that region of 
parameter space.  To facilitate that discussion, we explore 
this subsystem where analytic approximations are available, 
i.e., the limit where one Dirac mass eigenstate remains 
uncoupled from all of the other states.  Anticipating the 
following section, we decouple what is there $m_2$.  That 
is, we examine a two flavor system in which the Dirac mass 
eigenvalues are $m_1$ and $m_3$. 

It is useful to define:
\Bea
m_0^2  & = & m_1^2 \cos^2 \theta + m_3^2 \sin^2 \theta \label{1} \\
a & = &\frac{ \left(m_1^2 - m_3^2\right)\sin\theta \cos
\theta}{m_0\sqrt{2}} \label{2} \\
b & = & \frac{m_1m_3}{m_0} \label{3}
\Eea
and $c = \cos\theta$, $s = \sin\theta$.  Note the additional 
$1/\sqrt{2}$ factor in $a$.  These refer to the mass matrix
\Be
{\cal{M}}_{1} =
\left(\begin{array}{cccc}
0 & 0 & m_{1} & 0 \\
0 & 0 & 0 & m_{3} \\
m_{1} & 0 & Ms^{2} & Mcs \\
0 & m_{3} & Mcs & Mc^{2}
\end{array}
\right) 	\label{mass4}
\Ee
where $m_1, m_3$ are Dirac masses for the two neutrino flavors 
and $M$ is the single nonzero mass eigenvalue in the sterile 
sector. The angle $\theta$ defines the deviation of the direction 
in the sterile subspace of the eigenvector for this nonzero mass 
from one of the flavor axes defined by the Dirac mass eigenstates. 
Note that the structure in Eq.(\ref{mass4}) is equivalent to the 
assumption that the MNS\cite{MNS} analog of the CKM\cite{ckm} 
matrix for the quarks is the unit matrix. 

It is useful to transform ${\cal{M}}_{1}$ into the form
\Be
\cal{M} =
\left(\begin{array}{cccc}
m_0 & 0 & 0 & a \\
0 & -m_0 & 0 & -a \\
0 & 0 & 0 & b \\
a & -a & b & M
\end{array}
\right)
\Ee
in order to see that, to lowest order, the three small eigenvalues 
are $\pm m_0, 0$.  (Note the minus sign on the $a$ in the (2,4) and
(4,2) positions.) The matrix effecting the transformation ${\cal{M}} 
= \Omega^{\dag} {\cal{M}}_{1} \Omega$ is
\Be
\Omega =
m_{0}^{-1}\left(\begin{array}{cccc}
m_{1}s/\sqrt{2} & -m_{1}s/\sqrt{2} & m_{3}c & 0 \\
-m_{3}c/\sqrt{2} & m_{3}c/\sqrt{2} & m_{1}s & 0 \\
m_{0}s/\sqrt{2} & m_{0}s/\sqrt{2} & 0 & m_{0}c \\
-m_{0}c/\sqrt{2} & - m_{0}c/\sqrt{2} & 0 & m_{0}c
\end{array} \right)
\Ee

This suggests writing the characteristic equation as:
\Be
\mu \left(m_{0}^{2} - \mu^{2}\right)\mu(M - \mu) =
  2 \mu^{2} a^{2} - \left(m_{0}^{2} - \mu^{2}\right) b^{2}
\Ee
which is convenient for iterative solution in a series in $M^{-1}$.
The usual equation obtained directly from $\left| {\cal{M}}_{1} - 
\mu \right. ${\mtxfnt 1}$\left. \right| = 0$,
\Be
\mu^{4} - \mu^{3}M - \mu^{2}\left(m_{1}^{2} + m_{3}^{2}\right) + \mu
m_{0}^{2} M + m_{1}^{2} m_{3}^{2} = 0,
\Ee
is just the same equation.

The solutions to order $M^{-2}$ are 
\Bea
\mu_1 & = & m_0 - \frac{a^{2}}{M} -
\frac{a^{2}}{m_{0}M^{2}}\left(m_{0}^{2} - \frac{a^{2}}{2} -
b^{2}\right) \label{4solns1}  \\
\mu_2 & = & -m_0 - \frac{a^{2}}{M} +
\frac{a^{2}}{m_{0}M^{2}}\left(m_{0}^{2} - \frac{a^{2}}{2} -
b^{2}\right) \label{4solns2}  \\
\mu_3 & = & - \frac{b^2}{M} + {\cal O}(M^{-3}) \label{4solns3}  \\
\mu_4 & = & M +\frac{b^2}{M} +2 \frac{a^2}{M} +{\cal O}(M^{-3}) 
\Eea 

Notice that the eigenvalues sum to $M$ as they must and that 
the $\pm m_{0}$ eigenvalues are shifted in opposite directions 
at $O(M^{-2})$ but in the same direction at $O(M^{-1})$, which 
is a small amount for sufficiently large $M$. The latter shift 
is why these form a pseudo-Dirac pair rather than simply a 
Dirac bispinor.  Note also that $\mu_{3}$ and $\mu_{4}$, do 
not acquire $O(M^{-2})$ corrections; their next correction 
is at the next higher order.

Having obtained the eigenvalues, we now solve for the eigenvectors.
Since our interest is in the mixing in the active sector, it is useful
to carry this exercise out in the original representation, that of
${\cal{M}}_{1}$.  In this representation, we define the $i^{th}$ 
eigenvector as
\Be
\phi_i = \left( \begin{array}{c} \alpha_i \\ \beta_i  \\ \gamma_i 
\\ \delta_i \end{array} \right),
\Ee
where $\alpha_i$ and $\beta_i$ are the two active components and
$\gamma_i$ and $\delta_i$ are the two sterile components.

Picking three equations, we find
\Bea
-\mu_i \alpha_i + m_1 \gamma_i & = & 0  \nonumber \\
-\mu_i \beta_i + m_3 \delta_i & = & 0   \nonumber  \\
m_3 \beta_i  + M s c \gamma_i +(M c^2 - \mu_i) \delta_i & = & 0 \label{3eqs}
\Eea
A number of points are immediately clear from Eqs.(\ref{3eqs}): 
Since $\mu_{4} \sim M$, $\beta_{4}$ and $\alpha_{4}$ are small 
(${\cal O}(m_{D}/M)$) so the fourth eigenstate is almost entirely 
decoupled from the active sector.  Conversely, since $\mu_{3} 
\sim {\cal O}(m_{D}^2/M)$, $\gamma_{3}$ and $\delta_{3}$ are 
small (${\cal O}(m_{D}/M)$) so the third eigenstate resides almost 
entirely in the active sector. Finally, since $\mu_{1}$ and $\mu_{2}$ 
are of order ${\cal O}(m_{D})$, $\gamma_{1,2}$ and $\delta_{1,2}$ 
are comparable with $\beta_{1,2}$ and $\alpha_{1,2}$ so these two 
eigenstates are generally strongly mixed between the active and 
sterile sectors, i.e., they form a pseudo-Dirac pair. 

Substituting for $\gamma_i$ and $\delta_i$ gives an equation for the 
ratio
\Bea
\frac{\beta_i}{\alpha_i} & = & -\frac{M \mu_i s c}{[\mu_i (M c^2 -
\mu_i) + m_3^2]}  \nonumber \\  & = & -\frac{s c}{[c^2 -
\frac{\mu_i}{M} + \frac{m_3^2}{M \mu_i}]}. \label{ratio}
\Eea
Note that if either $s = 0$ or $c = 0$, one pair of states forms 
a purely Dirac bispinor and the other becomes the usual see-saw 
pair of Majorana states.

For the light mass eigenstates, the ratio $\beta_{i}/\alpha_{i}$ 
is a measure of mixing in the active sector. Solving Eq.(\ref{ratio}), 
we find that
\Be
| \frac{\beta_{1}}{\alpha_{1}} | \,\,\, =  
| \frac{\beta_{2}}{\alpha_{2}} | \,\,\, =
| \frac{\alpha_{3}}{\beta_{3}} | \,\,\, = 
\frac{m_{3}}{m_{1}} {\rm tan}(\theta), \label{lrgmx}
\Ee
where the last equality is correct to ${\cal O}(m_{D}^2/M^2)$ 
and the first two are correct to ${\cal O}(m_{D}/M)$. It is 
apparent that, in all three states, the mixing of the active 
components can be large simultaneously. 

Turning back to the amplitudes of the sterile components, 
we see from the first two lines of Eq.(\ref{3eqs}) that 
\Bea
\frac{\gamma_{i}}{\alpha_{i}} & = & \frac{\mu_{i}}{m_{1}} \NL
\frac{\delta_{i}}{\beta_{i}} & = & \frac{\mu_{i}}{m_{3}} .
\Eea	 
Hence, for a large range of values of ($m_{1}$, $m_{3}$, 
$\theta$), these ratios are ${\cal O}(1)$ for $i=1$ and 
$2$, which is, of course, characteristic of a pseudo-Dirac 
pair. As long as $M$ is large, $\gamma$, $\delta$ are small 
for $i=3$ since $\mu_{3} \sim 0$, and huge for $i=4$ since 
$\mu_{4} \sim M$. This reiterates the fact that the massive 
sterile state is quite decoupled, while the light sterile 
can be strongly coupled into the active states and the 
pseudo-Dirac states significantly mixed across all four 
components. 

%\subsection{Two Flavor Conclusion}

Thus, in this simple, two flavor model, we have demonstrated 
that a misalignment of the direction vector for the heavy 
sterile mass with the axes determined by the Dirac mass 
eigenstates necessarily induces mixing in the active sector 
for all of the light Majorana mass eigenstates, even with a
unit MNS matrix for the Dirac mass matrix. This point has 
been raised previously in Ref.\cite{rabi} in a different context. 

Moreover, large mixing of active states results over a region 
of the $(m_1, m_3, \theta)$ parameter space where the mass ratio 
and the deviation angle of the sterile components from flavor 
alignment approximately compensate, i.e., near the line determined 
by setting the rightmost quantity in Eq.(\ref{lrgmx}) to one. 
Mixing in the Dirac sector by an MNS matrix should not alter the 
general feature of achieving large mixing "naturally". 

Finally, we note explicitly the difference in oscillation structure 
between this $4\times4$ neutrino mass matrix and the $2\times2$ 
Majorana (or Dirac) mixing usually applied to interpret experiments. 
As shown by Eqs.(\ref{3}) or Eqs.(\ref{4solns1}, \ref{4solns2}, 
\ref{4solns3}), instead of one mass difference, here 

\pagebreak 

\noindent there are at least two 
independent mass differences, even in the limit of large sterile 
mass ($M$). Thus, simple two channel analyses are not guaranteed 
to extract the true physical oscillation parameters from experimental 
results. This problem is exacerbated in the $6\times6$ case that 
we discuss in the next Section, in which at least four independent 
mass differences appear where it has been conventionally assumed that 
there can only be two. 

\section{General mass matrix}\label{genmass}

The flavor basis for the active neutrinos and the pairing to 
sterile components defined by the (generally not diagonal) 
Dirac mass matrix could be used to specify the basis for the 
sterile neutrino mass matrix, $M_S$. Instead we take the basis 
in the $3\times 3$ sterile subspace to allow the convention 
described below. This implies a corresponding transformation 
of the Dirac mass matrix, which is irrelevant at present 
since the entries in that matrix are totally unknown. 

We define our convention for the choice of axes in the $3\times 3$ 
sterile subspace as follows. Denote the nonzero mass eigenvalue 
of the rank $1$ by $M$ and choose its eigenvector initially 
in the third direction. Then rotate this vector, first by 
an angle of $\theta$ in the $1-3$ plane and then by $\phi$ 
in the $1-2$ plane. The rotation is chosen so that the 
Dirac mass matrix which couples the active and sterile 
neutrinos becomes diagonal, i.e., the basis is defined by 
Dirac eigenstates. This produces a $3 \times 3$ mass matrix 
in the sterile sector denoted by
\Be
M_S = M \left[ 
\begin{array}{ccc}
\cos^2 \phi \sin^2 \theta  &  \cos \phi \sin \phi \sin^2 \theta  &
\cos \phi \sin \theta \cos \theta  \\  \cos \phi \sin \phi \sin^2 \theta & 
\sin^2 \phi \sin^2 \theta  &  \sin \phi \sin \theta \cos \theta  \\ 
\cos \phi \sin \theta \cos \theta & \sin \phi \sin \theta \cos \theta & 
\cos^2 \theta   \end{array}  \right]. 
\Ee

In this representation, the Dirac mass matrix is diagonal 
by construction
\Be
m_D = \left[ \begin{array}{ccc}
m_1   &  0  &  0  \\  0  &  m_2  &  0  \\  0  &  0  &  m_3
\end{array} \right].  
\Ee

Note that there are special cases.  For $\theta = 0$ and any 
value for $\phi$,
\Be
M_S = \left[ \begin{array}{ccc}
0 & 0 & 0 \\ 0 & 0 & 0 \\ 0 & 0 & M \end{array} \right]  .
\Ee
For $\theta = \pi / 2$ and $\phi = 0$, 
\Be
M_S = \left[ \begin{array}{ccc} 
M & 0 & 0 \\ 0 & 0 & 0 \\ 0 & 0 & 0 \end{array} \right], 
\Ee
and, for $\theta = \pi / 2$ and $\phi = \pi / 2$, 
\Be
M_S = \left[ \begin{array}{ccc} 0 & 0 & 0 \\ 0 & M & 0 \\ 0 & 0 & 0
\end{array} \right]. 
\Ee
These are equivalent under interchanges of the definition 
of the third axis. 

The $6 \times 6$ submatrix\cite{fn2} of the full $12 \times 
12$ is, in block form, 
\Be
{\cal M} = \left[ \begin{array}{cc} 0 & m_D \\ m_D & 
M_S \end{array} \right].   
\Ee
Since we are ignoring CP violation here, no adjoints or 
complex conjugations of the mass matrices appear. 

Note that, in the chiral representation, the full $12 \times 12$ matrix is
\Be
\left[ \begin{array}{cc} 0 & {\cal M} \\ 
{\cal M} & 0 \end{array} \right]. \nonumber
\Ee
Thus the full set of eigenvalues will be $\pm$ the eigenvalues of
${\cal M}$.  Where it matters for some analysis we keep track of 
the signs of the eigenvalues, however for most results we present 
positive mass eigenvalues.

After some algebra, we obtain the secular equation 
\Bea
0 & = & \lambda^6 -M \lambda^5 -(m_1^2 + m_2^2 + m_3^2) \lambda^4 \nonumber \\
& & + M [m_3^2 \sin^2 \theta + m_2^2 (\sin^2 \theta \cos^2 \phi + \cos^2 \theta)] 
\lambda^3 \nonumber \\ & & + (m_1^2 m_2^2 + m_2^2 m_3^2 + m_3^2 m_1^2) 
\lambda^2 \\
& & - M (m_1^2 m_2^2 \cos^2 \theta + m_2^2 m_3^2 \cos^2 \phi \sin^2 \theta 
\nonumber \\
  &  & + m_3^2 m_1^2 \sin^2 \phi \sin^2 \theta) \lambda \nonumber \\
  &  & - m_1^2 m_2^2 m_3^2. \nonumber
\Eea  

\pagebreak

This may be rewritten as 
\Bea
 0 & = & (\lambda^2 - m_1^2) (\lambda^2 - m_2^2) 
(\lambda^2 - m_3^2) \nonumber \\ &   &    
- \lambda M \left( \lambda^4 -\left[ m_3^2 \sin^2 \theta 
+ m_2^2 (\sin^2 \theta \cos^2 \phi + \cos^2 \theta ) \right. \right. \nonumber \\
&  & \left. \left. + m_1^2 ( \sin^2 \theta \sin^2 \phi + \cos^2 \theta) 
\right] \lambda^2 \right. \\
&  & \left. +m_1^2 m_2^2 \cos^2 \theta + m_2^2 m_3^2 \sin^2 \theta 
cos^2 \phi  \right. \nonumber \\
&  & \left. + m_3^2 m_1^2 \sin^2 \theta \sin^2 \phi \right). \nonumber
\Eea

The special cases follow directly.  For $\theta = 0$, we find
\Be
(\lambda^2 - m_1^2) (\lambda^2 - m_2^2)
(\lambda^2 - M \lambda - m_3^2) = 0, 
\Ee
for $\theta = \pi / 2$ and $ \phi = 0$
\Be
(\lambda^2 - m_2^2) (\lambda^2 - m_3^2)
(\lambda^2 - M \lambda - m_1^2) = 0, 
\Ee
and for $\theta = \pi / 2$ and $\phi = \pi / 2$
\Be
(\lambda^2 - m_3^2) (\lambda^2 - m_1^2)
(\lambda^2 - M \lambda - m_2^2) = 0. 
\Ee

If $m_1^2 = m_2^2 = m_3^2 = m^2$, then we find
\Be
(\lambda^2 - m^2)^2 (\lambda^2 - M \lambda - m^2) = 0. 
\Ee

Due to the wide range of possibilities inherent in the system, 
it is useful to examine specific numerical examples.  For the
immediate exercise, we have picked the following parameters:
$m_1 = 1$, $m_2 = 2$, $m_3 = 3$ and $M = 1000$. The relatively 
small value of $M$ is chosen so that the splittings are not so 
tiny as to be difficult to discern. 

For this choice, the eigenvalues have a definite pattern 
for all values of $\theta$ and $\phi$.  There are two very 
close pairs, with mass eigenvalues between $1$ and $3$.  
There is one very small eigenvalue, of order $10^{-3}$ 
reflecting the ratio of $m_D$ to $M$, and one large eigenvalue 
of order $10^{3}$ (i.e., of order $M$).  Treating the last 
two as a pair despite their disparity in mass allows us to 
present results in tabular form, one for each pair, for sets 
of angles $\theta , \phi = \pi / 8, \pi / 4, 3 \pi / 8 $.

First, for the lower mass close pair, we have
\Be
\begin{array}{lccc}
\theta \backslash \phi & \pi /8  &  \pi /4  & 3 \pi / 8  \\
             &         &          &            \\
\pi / 8      & 1.398125 & 1.230175 & 1.068477   \\
             & 1.394934 & 1.228025 & 1.067688    \\
             &          &          &            \\
\pi / 4      & 1.809478 & 1.478863 & 1.151936   \\
             & 1.808183 & 1.477134 & 1.150941   \\
             &          &          &             \\
3 \pi / 8    & 1.877166 & 1.562977 & 1.18999    \\
             & 1.876742 & 1.561911 & 1.189146  \end{array}
\Ee

Then, for the next mass pair with close eigenvalues, 
we find
\Be
\begin{array}{lccc}
\theta \backslash \phi & \pi / 8  &  \pi / 4 & 3 \pi / 8   \\
             &          &          &             \\
\pi / 8      & 2.038992 & 2.107688 & 2.158044    \\
             & 2.038729 & 2.107156 & 2.157407    \\
             &          &          &              \\
\pi / 4      & 2.347974 & 2.46348  & 2.529128    \\
             & 2.346047 & 2.462176 & 2.52809     \\
             &          &          &              \\
3 \pi / 8    & 2.816525 & 2.847539 & 2.868607    \\
             & 2.815691 & 2.846972 & 2.868186   
\end{array} 
\Ee

\pagebreak 

Finally, even though it does not directly impact the 
argument, we display the remaining pair in order to 
present a complete set.  
\Be
\begin{array}{lccc}
\theta \backslash \phi & \pi / 8  &  \pi / 4 & 3 \pi / 8   \\
             &          &          &             \\
\pi / 8      & 1000.008 & 1000.008 & 1000.008    \\
             & 0.00444  & 0.005366 & 0.006778    \\
             &          &          &              \\
\pi / 4      & 1000.005 & 1000.006 & 1000.006    \\
             & 0.001997 & 0.002717 & 0.004248    \\
             &          &          &              \\
3 \pi / 8    & 1000.003 & 1000.003 & 1000.004    \\
             & 0.001289 & 0.001819 & 0.003092   
\end{array} 
\Ee

\section{Two nearly degenerate pseudo-Dirac pairs}\label{2pds}

Applying the techniques of the last section, we find the angle 
$\theta$ such that $m_2$ and the eigenvalue for the pseudo-Dirac 
pair above, $m_0$, are approximately degenerate.  We then vary  
$\phi$ away from $0$ and display the eigenfunctions.  To 
illustrate the general nature of the result, we have changed 
the Dirac masses from the even spacing used above.

In Table I, the Dirac masses are taken to be $m_1 =  1$, $m_2 
=  1.1$, and  $m_3 =  3$.  The value for $m_2$ has been changed 
from above so that we can demonstrate that small angles in the 
sterile sector can lead to large mixing in the active sector.  
Again, in order to display the structure of the spectrum, we have 
chosen $M = 1000$, rather than a larger value, expected to be more 
realistic, but which would suppress the difference scale between 
the pairs. The angles are given in degrees.

Table I represents only a small part of the available parameter 
space; the values of the angles are chosen to display some 
interesting possible features. First, $\theta$ has been chosen so 
that, at $\phi = 0$, the Dirac pair at $m_2$ is bracketed by the 
pseudo-Dirac pair.  Such a value of $\theta$ exists for any pattern 
of the Dirac masses.  Then, for small values of $\phi$, there are 
always two nearly degenerate pseudo-Dirac pairs.

Note that, for $\phi = 0$, there is no mixing between the 
field labelled by $2$ and the remaining fields, while for 
the next entry at $\phi = 2.25$ degrees there is considerable 
mixing.  That mixing increases with $\phi$ as the difference 
bewteen the eigenvalues increases.  The pattern described by 
the centroids of the pseudo-Dirac pairs is fixed by the angles 
$\theta$ and $\phi$.  If $M$ is increased, that pattern hardly  
changes.  The primary effect of increasing $M$, consistent 
with the analysis in Sec.\ref{2flavor}, is to decrease the 
separation of the two members of each pseudo-Dirac pair while 
producing the usual see-saw behavior for the remaining pair. 
Thus, tiny differences in mass between masses that are not 
especially small themselves, are, in the usual sense of the 
term, natural in this approach. 

The implication for oscillation phenomena is clear.  A given 
weak interaction produces an active flavor eigenstate which 
is some linear combination of the three active components 
listed in Table I.  That then translates into a linear 
combination of the six mass eigenstates.  From Table I, it 
is clear that the involvement of the heavy Majorana see-saw 
state is minimal, so the system effectively consists of the 
light Majorana see-saw state and the four Majorana states 
arising from the two pseudo-Dirac pairs. These five states 
include all three active neutrinos, generating a natural 
3+2 scenario. 

Since these five mass eigenstates have both active and 
sterile components, the subsequent time evolution will 
involve both flavor changing oscillations and oscillation 
into (and back out of) the sterile sector. This can lead 
to very complex oscillation patterns, as there are $10$ 
mass differences, $4$ of which are independent. A specific 
example is discussed in the next section. 

Finally, inspection of the column labelled ``1active'' for 
$\phi = 2.25$ or $\phi = 4.5$, for example, shows that 
the presence of a rank $1$ sterile mass matrix can 
seriously change any mixing pattern of the MNS type\cite{MNS}, 
from that which would have obtained with purely Dirac neutrinos.

\section{example}\label{example}

In Figs.1 to 3, we plot the oscillation patterns that appear for 
the parameters set by the second entry in Table I.  Fig.1 gives 
an overview of the case where an active neutrino (labelled $1$) 
is produced initially. The plot is given versus $L/E$, where $L$ 
is the distance from the neutrino source and $E$ is the energy 
(bin) of the neutrino observed. (As shown in the Appendix, $L/p$,  
where $p$ is the momentum of the neutrino, might well be the 
more correct variable to use, but the difference is certainly 
irrelevant in all conceivable neutrino experiments.) 

The (compressed scale) Fig.1 shows rapid oscillations between 
active neutrinos $1$ and $3$ with a later appearance of the 
active neutrino $2$. Note the large mixing among all three 
channels of active neutrinos. The mixing to sterile neutrinos 
is large also, but occurs on a much larger $L/E$ scale, 
corresponding to the much smaller mass difference (approximate 
degeneracy) of the pseudo-Dirac pairs. 

Fig.2 extracts from Fig.1 the appearance of active neutrino 
$2$ at small $L/E$ (short baseline experiments). Clearly, 
attempting to fit this highly nonsinusoidal behavior with two 
channel sinusoidal mixing will generally not yield physical 
mixing parameters in good agreement with the actual three 
channel case. 

Fig.3 emphasizes how such an error may be magnified by limited 
resolution in an experiment. The heavy black curve is the average 
over 100 $L/E$ units of the probability for finding the initial 
neutrino flavor. It approximates the shape of the {\bf envelope} 
of the high frequency oscillations. A two channel analysis would 
clearly find a small difference between the squared masses for 
the mixing from active neutrino $1$ to active neutrino $3$ 
despite the obviously larger value demonstrated by the rapid 
oscillation cycles. A similar conclusion follows from the 
cycle-averaged curve for appearance of active neutrino $3$. 
Finally, one would be tempted to conclude that the mixing to 
active neutrino $2$ is small or negligible, when in fact it 
is about as large as any other mixing in the full case. 

Labelling active neutrino $1$ as the muon neutrino, active neutrino 
$3$ as the tau neutrino and active neutrino $2$ as the electron 
neutrino illustrates our concerns about the strong conclusions 
drawn from atmospheric and accelerator neutrino experiments by 
means of two channel mixing analyses. Similar concerns\cite{tw} 
have also been raised in the literature previously. 

\section{Rank $2$}

We have not discussed the case of rank $2$ matrices explicitly,
although the pattern is obvious.  In such a case, there would 
be two see-saw pairs and one pseudo-Dirac pair, leading to 
three active and one sterile light neutrino.  While this pattern 
has been analyzed in the literature, we do not find any compelling 
pattern for it in the sterile sector.  Furthermore, the current 
consistency of all neutrino oscillation data can be accomodated 
much more easily (and perhaps only, as indicated by Ref.\cite{Sorel},) 
in the rank $1$ case discussed in this paper. Therefore we do  
not discuss rank $2$ at this time.

\section{Conclusions}

We have considered here the effects on neutrinos in the SM of the 
recurrently successful and conventional constraint of quark-lepton 
symmetry, namely, the existence of six independent Weyl spinor 
fields of neutrinos, three corresponding to active and three 
corresponding to sterile neutrinos. In the now venerable see-saw 
approach, the latter three effectively disappear from the excitation 
spectrum, leaving small Majorana masses for the active states as a 
residuum. We have examined the effect on this system of a rank less 
than three character of the $3 \times 3$ mass matrix in the sterile 
sector and studied the rank $1$ case, in particular. 

In the rank $1$ case on which we have focused, we find that 
the neutrino fields naturally form into two pseudo-Dirac 
pairs, leaving only one almost pure Majorana active neutrino 
and one conventionally very heavy sterile Majorana neutrino. 
More importantly, we also find a naturally strong mixing 
between the active and sterile parts of the two pseudo-Dirac 
pairs.  Further, we find that this can easily affect the 
mixing between active neutrinos even if it is otherwise small. 
That is, even if the Dirac mass matrix induced mixing analogous 
to what is known to occur in the quark sector is small or absent, 
mixing between active neutrinos can develop with large values. 
In a two flavor case, we demonstrated analytically that this 
strong mixing can develop over a wide range of parameters.

We have chosen a limited relative value of the sterile neutrino 
mass scale, $M$, that allows for easy discernment of the nature 
of the effects. It should be noted, however, that the primary 
effect of increasing $M$ is to decrease the separation of the 
two members of each pseudo-Dirac pair while producing the usual 
see-saw behavior for the remaining pair. Thus, tiny differences 
in mass between masses that are not especially small themselves 
are, in the usual sense of the term, natural in this approach. 
This is contrary to the general expectation that the small mass 
differences responsible for the observed neutrino oscillation 
phenomena presage small absolute masses for all of the neutrinos. 
Furthermore, increasing the value of $M$ without altering the 
Dirac masses retains the features and scales of the oscillations 
essentially unchanged; the only significant change is that the 
appearance of the sterile components is delayed to even greater 
values of $L/E$. 

The features described above are most easily discerned in the 
case when the Dirac mass terms for the neutrinos are well separated 
in value. It remains conceivable that, if their differences are 
small for some other reason, then the splitting between the pseudo-Dirac 
pairs may be larger than that between flavors. In this case, it 
is still true that large flavor mixing is naturally induced.

Finally, we presented a specific model which raises concerns 
about the conclusions drawn from analyses of neutrino oscillation 
experiments in terms of two channel mixing: Such analyses may 
be misleading regarding the true values of physical parameters. 
After the completion of this work, we learned of papers\cite{ernst} 
which have independently suggested that such a concern may well be 
justified. 

\vspace*{-0.1in}
\section{Acknowledgments}
\vspace*{-0.1in}
This research is supported in part by the Department of Energy under
contract W-7405-ENG-36, in part by the National Science Foundation 
under NSF Grant \# PHY0099385 and in part by the Australian Research 
Council.

\newpage

\section{APPENDIX: Two Mass Eigenstate Oscillations}

There has been some confusion and discussion in the literature 
regarding the space and time dependence of neutrino oscillations. 
We briefly present here an argument in the rest frame of a state 
of a given flavor that demonstrates the time dependence unequivocally. 
By boosting the observer instead of the state, we demonstrate the 
equivalence of the usual $L/E_{\nu} \approx L/p_\nu$ dependence, 
derived in several ways, to the time dependence in the rest frame. 
Hence the figures in the main sections of this paper can be viewed 
as either variation with $L/E_{\nu}$, $L/p_\nu$ or time. 

We begin with a flavor eigenstate composed of two different mass 
eigenstate contributions, in their common rest frame. Let
\Bea
c \equiv cos(\theta) & ; & s \equiv sin(\theta) \NL
|\nu_{f}> & = & \;\;\, c \, |\nu_1> + s \, |\nu_2> \NL
|\nu_{g}> & = & -s \, |\nu_1> + c \, |\nu_2> 
\Eea 
to fix conventions for neutrino flavors $f$ and $g$ composed 
in two channel mixing from neutrino mass eigenstates $\nu_1$ 
with mass $m_1$ and $\nu_2$ with mass $m_2$. The time evolution 
of the state initially in flavor $f$ in this common rest frame 
is given by
\Bea
& & |\nu(t)> = e^{-\imath {\bf H} t} \, |\nu(0) \equiv \nu_{f}> \NL
& = & c \, e^{-\imath m_1 t} |\nu_1> + s \, e^{-\imath m_2 t} |\nu_2> 
\Eea 
The probability of appearance of ${\nu}_{g}$ from the ${\nu}_{f}$ 
source is given by
\Bea
& & |<\nu_{g}|\nu(t)>|^2  = |e^{-\imath\frac{(m_1+m_2)t}{2}}|^2 \NL 
& \times & c^{2}s^{2} \, | -e^{\imath\frac{(m_2-m_1)t}{2}} + 
e^{-\imath\frac{(m_2-m_1)t}{2}}|^2 \NL
& = & sin^{2}(2\theta) \, sin^{2}(\frac{\Delta m\, t}{2}) 
\Eea 
Viewed from a frame moving past these states with velocity $\beta$, 
the relation between position $L$ in the moving frame and the 
time is affected both by the velocity and by time dilatation, i.e., 
\Be
t = L/[{\beta \gamma}] \sim L m_{{av}}/p_{{\nu}}  , 
\Ee
where we choose the motion of the frame to be consistent with 
the ratio of the average neutrino mass and mean neutrino momentum. 
Hence
\Be
|<\nu_{g}|\nu(t)>|^2 \sim sin(2\theta) \, sin^2({\Delta m^2} 
L/[4 p_{{\nu}}])
\Ee
As usual, the units are determined by the relation 
\Be
{1.27} \frac{\Delta m^2}{eV^2} \frac{L_{osc}}{km} 
\frac{GeV/c}{p_{\nu}} = \pi
\Ee
consistent with all conventional analyses and expectations. 

It should be noted that different energy (mass) eigenstates do 
not interefere with each other. The effect derives entirely 
from the independent phase advance of the individual states and 
their translation relative to the laboratory rest frame. 

\newpage

\noindent TABLE I: Eigenmasses for various values of $\theta$ and 
$\phi$ for cases of two approximately degenerate pseudo-Dirac pairs.
\begin{verbatim}
__________________________________________________________________________
\end{verbatim}
$\theta =  9.324078$,   $\phi =  0$

\begin{verbatim}
      mass     1active    2active   3active  1sterile  2sterile  3sterile

   1.099328   0.635032   0.000000 -0.310533  0.698108  0.000000 -0.113793
   1.100680  -0.633620   0.000000  0.314383  0.697413  0.000000 -0.115345
   1.100000   0.000000   0.707107  0.000000  0.000000  0.707107  0.000000
   1.100000   0.000000  -0.707107  0.000000  0.000000  0.707107  0.000000
   0.007438   0.441883   0.000000  0.897064 -0.003287  0.000000 -0.002224
1000.008789   0.000162   0.000000  0.002960  0.162017  0.000000  0.986784
\end{verbatim}

$\theta =  9.324078$,   $\phi =  2.25$

\begin{verbatim}
      mass     1active    2active    3active  1sterile  2sterile  3sterile

   1.095953   0.479130  -0.468214  -0.225940  0.525106 -0.466489 -0.082539
   1.096608   0.437964  -0.514829  -0.208027 -0.480274  0.513243  0.076041
   1.103359   0.416946   0.529767  -0.212981  0.460041  0.531383 -0.078333
   1.104056  -0.458049  -0.484588   0.235669  0.505710  0.486376 -0.086730
   0.007438   0.441553   0.015769   0.897088 -0.003285 -0.000109 -0.002224
1000.008789   0.000162   0.000007   0.002960  0.161892  0.006361  0.986784
\end{verbatim}

$\theta =  9.324078$,   $\phi =  4.5$

\begin{verbatim}
      mass     1active    2active    3active  1sterile  2sterile  3sterile

   1.092254   0.479875  -0.471453  -0.217491  0.524155 -0.468127 -0.079183
   1.092888  -0.458763   0.495815   0.209390  0.501371 -0.492614 -0.076279
   1.107010   0.416602   0.526536  -0.221472  0.461189  0.529886 -0.081725
   1.107726   0.437718   0.503654  -0.234323 -0.484866 -0.507196  0.086521
   0.007439   0.440571   0.031517   0.897156 -0.003273 -0.000217 -0.002226
1000.008789   0.000162   0.000014   0.002960  0.161517  0.012712  0.986784
\end{verbatim}

$\theta =  9.324078$   $\phi =  22.5$

\begin{verbatim}
      mass     1active    2active    3active  1sterile  2sterile  3sterile

   1.062925   0.550356  -0.405921  -0.179528  0.584987 -0.392239 -0.063608
   1.063381   0.546548  -0.411257  -0.179609 -0.581185  0.397574  0.063663
   1.134871   0.337840   0.568726  -0.249457  0.383405  0.586755 -0.094367
   1.135731   0.341702   0.564710  -0.254038 -0.388074 -0.583058  0.096172
   0.007475   0.409265   0.154109   0.899298 -0.003058 -0.001048 -0.002241
1000.008789   0.000150   0.000068   0.002960  0.149684  0.062001  0.986784
\end{verbatim}

$\theta =  9.324078$   $\phi =  45$

\begin{verbatim}
      mass     1active    2active    3active  1sterile  2sterile  3sterile

   1.030458   0.632073  -0.290233  -0.127244  0.651329 -0.271878 -0.043708
   1.030692  -0.630801   0.292859   0.127989  0.650162 -0.274406 -0.043972
   1.163620   0.226485   0.612428  -0.270973  0.263544  0.647849 -0.105102
   1.164612   0.227955   0.610618  -0.274587 -0.265472 -0.646488  0.106595
   0.007566   0.315141   0.286490   0.904762 -0.002384 -0.001969 -0.002282
1000.008789   0.000115   0.000126   0.002960  0.114563  0.114563  0.986784
________________________________________________________________________
\end{verbatim}

\begin{figure*}[h] 
\includegraphics[height=4.0in]{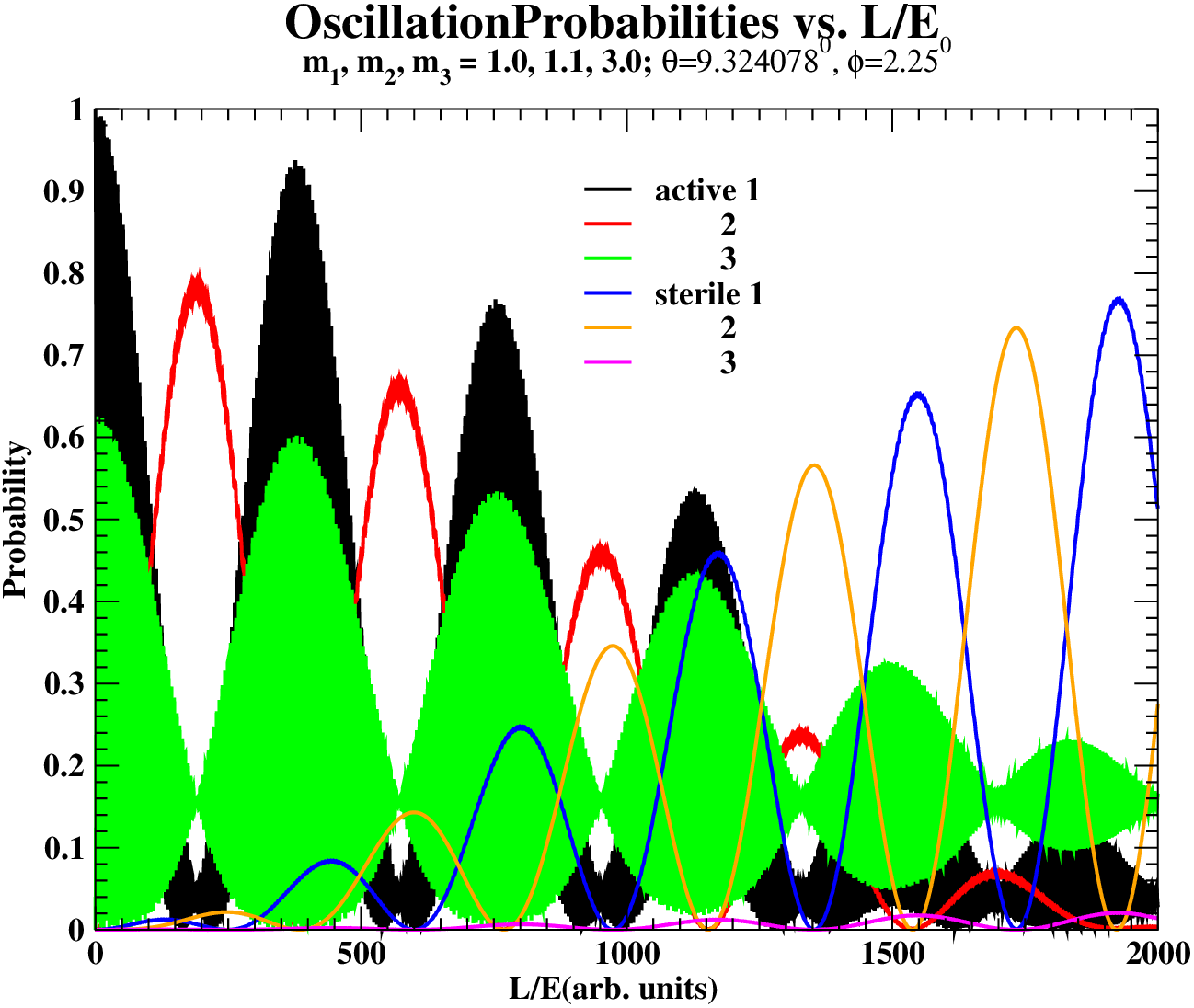} 
\caption{ Oscillations for all six channels commencing 
from one active flavor with appearance probabilities for 
the two other actives and the steriles using the second 
set of angle parameters in Table I.} 
\label{2K} 
\end{figure*} 

\begin{figure*}[h] 
\includegraphics[height=4.0in]{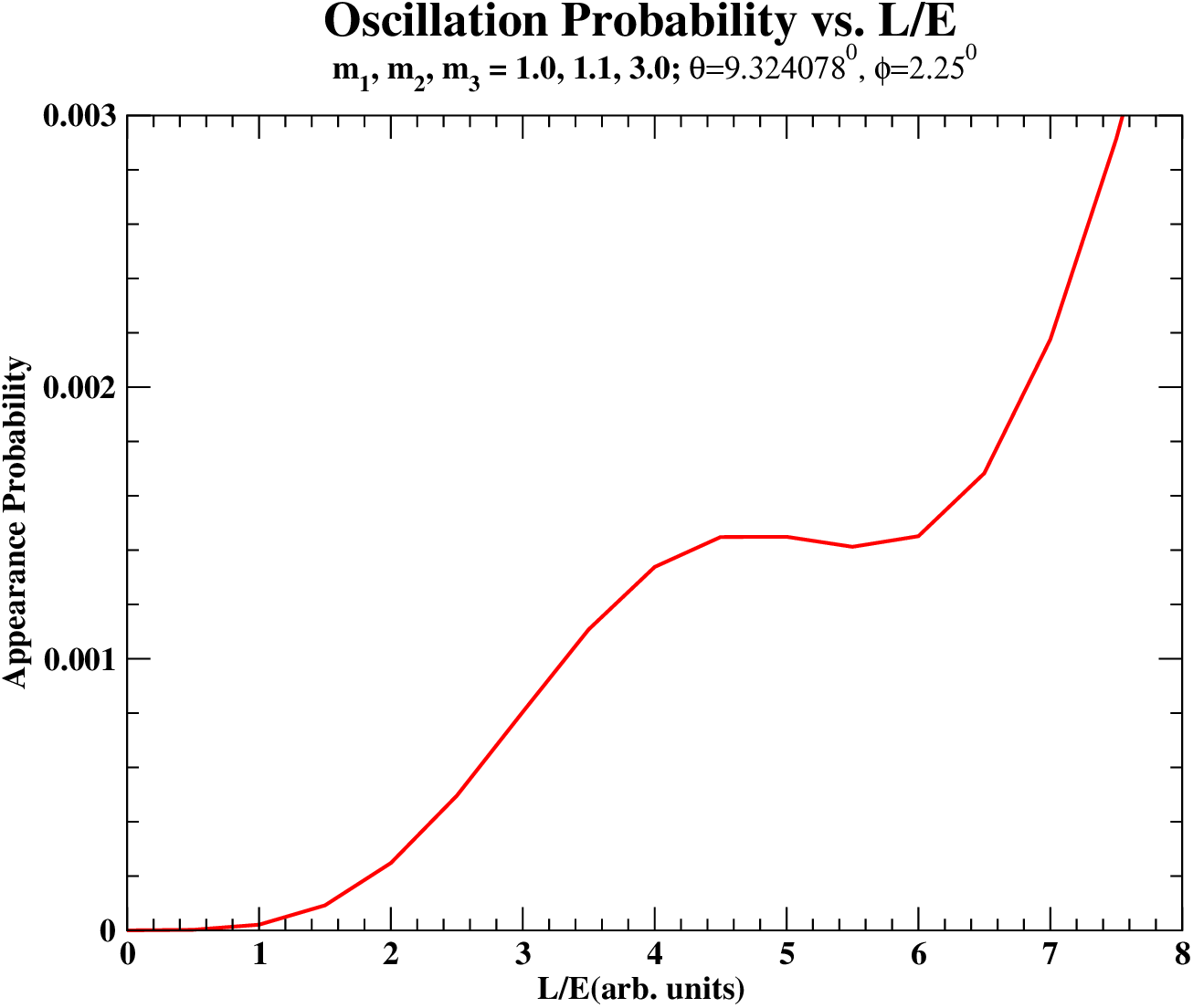} 
\caption{ Appearance probability for one neutrino 
flavor using the second set of angle parameters 
in Table I. The arbitrary units of the abscissa may 
be viewed as time in the rest frame or $L/E$ in the 
laboratory. See Appendix.} 
\label{appear} 
\end{figure*} 

\begin{figure*}[h] 
\includegraphics[height=4.0in]{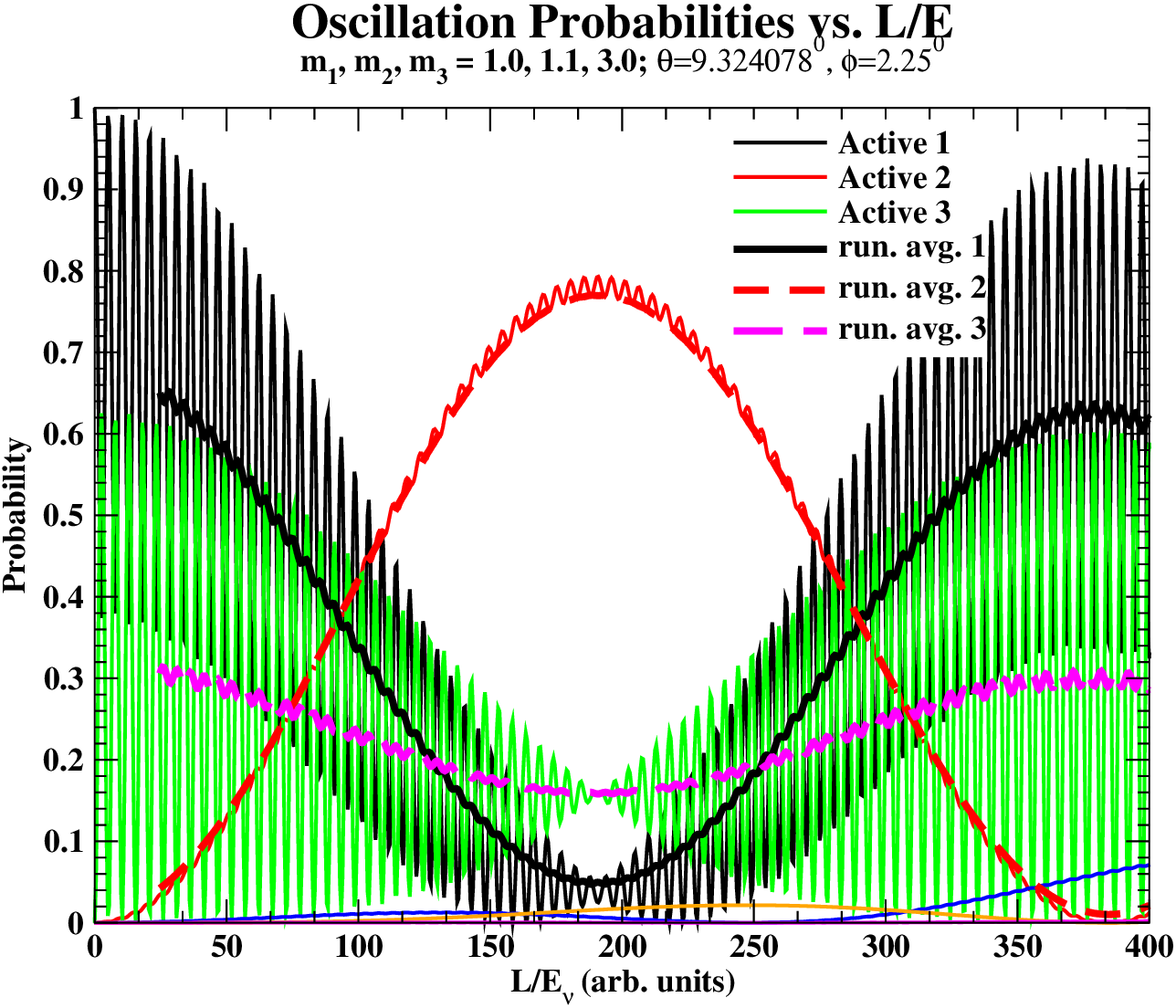} 
\caption{ Effects of limited resolution in $L/E$. 
Running averages over 100 units have been taken for 
each of the active curves. (The curves for the sterile 
neutrinos remain as in Fig.1.)} 
\label{runavs} 
\end{figure*} 


\begin{thebibliography}{99}
\bibitem{fn1} {In much of the literature, the sterile space is 
	referred to as ``Right-handed'' (or just R) and the 
	active space as ``Left-handed'' (or just L), which 
	follows from the behavior of the components of a Dirac 
	neutrino where the neutrino is defined as that neutral 
	lepton emitted along with a positively charged lepton. 
	Since we are dealing with mass matrices, which necessarily 
	all couple left-handed representations to right-handed 
	representations, we choose to refer to the Weyl neutrino 
	representations as active and sterile. ``Sterile'' refers 
	only to the SM; these neutrino states may have non-SM 
	interactions.}
\end{thebibliography}
\end{document}